\documentclass{svjour3}
\usepackage[applemac]{inputenc}
\usepackage{footmisc}

\usepackage{amscd,bbm}
\usepackage{natbib}
\usepackage{mathrsfs}
\usepackage{hyperref}
\usepackage[all]{hypcap}
\usepackage{braket}
\journalname{Arxiv}
\begin{document}
\title{The measurement problem and scientific realism\footnote{The first version of this paper was written in 2019.}}
\institute{Jorge Manero: Universidad Nacional Autónoma de México, Instituto de Investigaciones Filosóficas, Universidad 3000, Ciudad Universitaria, 04510, Mexico City, Mexico. Email: jorge.manero@filosoficas.unam.mx; ORCID: 0000000345885132.  \\ Juan Guzmán: Faculty of Science, Investigación Científica, C.U., Coyoacán, 04510, Mexico City, Mexico; ORCID: 0000-0001-7142-284X}
\date{}
\author{Jorge Manero \and Juan Guzmán}
\maketitle
\begin{abstract}
Many attempts have been made to characterise and solve the infamous measurement problem of quantum mechanics by advocating, implicitly or explicitly, different realist perspectives. As a result, we are still uncertain where this problem and its corresponding solution are to be located in the realism-antirealism debate. On the basis of a well-known characterisation of scientific realism, this paper intends to fill this gap by arguing that the quantum description of the processes involved in typical measurements is problematic from a minimal realist point of view, known as \emph{semantic realism}. 
\end{abstract}
\section{Introduction}
It is well known that the feasibility of interpreting \emph{quantum mechanics} (QM) realistically is controversial, in part because this theory faces numerous conceptual issues that cannot be addressed without directly participating in the realism-antirealism debate. Among these issues we find the \emph{measurement problem}. Although alternative quantum theories and interpretations have been suggested in order to solve this problem, we do not yet have a definite picture of what exactly might reconcile these alternatives with scientific realism. For example, beyond the standard way to formulate QM, including \emph{unitary quantum mechanics} (UQM) and \emph{standard quantum mechanics} (SQM), some of these alternative  interpretations (to be addressed in this contribution) comprise \emph{Frauchiger-Renner formulation} (FRQM), some \emph{non-collapse interpretations} (NSQM), the \emph{Bohmian approach} (BQM) and \emph{objective collapse interpretations} (GRW). By UQM we shall mean the unitary theory that excludes from SQM (the one usually reviewed in textbooks) any reference to the collapse postulate but retains the completeness assumption with respect to the wave function. By FRQM we shall mean the theory associated with a particular interpretation of \citep{neumann} that, to the best of our knowledge, forms the basis of the extended Wigner's friend experiment proposed by \citep{frauchiger}. Finally, by NSQM we shall mean the set of non-collapse theories that are both unitary and include some interpretation of QM, such as the many-worlds interpretation, relational quantum mechanics, the modal interpretation and the consistent histories approach.\footnote{We shall assume the reader is acquainted with BQM and GRW.}

Through a critical revision of how these quantum theories and interpretations have characterised the measurement problem and/or supposedly solved it, our main contribution consists in understanding in what precise realist sense this problem associated with QM is a problem for quantum philosophers. We seek to demonstrate that, to put it crudely, \emph{the measurement problem is a problem for semantic realism}, which is a necessary but not sufficient condition for fully fledged scientific realism. Considering this analysis, we shall support the claim that UQM, SQM, FRQM and NSQM entail serious concerns regarding objectivity, understanding, and predictive reliability in our theoretical dealings with phenomena, all of which can be overcome by elaborating, clarifying and extending the objective correspondence that is hypothesised to hold between these theories and the world they describe.\footnote{In contrast, advocates of \emph{perspectivalism} or \emph{rainforest realism} believe that this problem can be \emph{dissolved} by abandoning some deeply rooted epistemological and metaphysical commitments incompatible with QM. Although these approaches deserve special attention, they shall lie beyond the scope of this paper.} In so doing, we will be able to identify precisely where this problem and its corresponding solution are to be located in the realism-antirealism debate. 

Our methodological procedure goes as follows: in Sections \ref{sec2} and \ref{sec3} we lay down an appropriate characterisation of scientific realism based on physics; in Section \ref{sec4} we proceed to describe the measurement process; in Section \ref{sec5} we support the claim that the measurement problem is a semantic realist struggle by identifying the underlying realist criteria that are at stake in QM; finally, in Section \ref{sec6}, we write some concluding remarks. 
\section{The scientific realism triad}\label{sec2}
Following the standard literature in the field \citep{van,psillos,real}\footnote{NB: we focus only on the way these authors define scientific realism, regardless of whether or not they endorse it.}, physics-based scientific realism may be generally defined in terms of the conjunction of three necessary components. In addition to the well-known mind-independent condition (i.e., a necessary condition for fully fledged scientific realism that shall be taken for granted), the semantic component (hereinafter \emph{semantic realism}) is, put simply, the view concerning how physical theories are to be interpreted and what they are supposed to be saying without assuming that they are correct, whilst the epistemic component (hereinafter \emph{epistemic realism}) states, in addition to this semantic correspondence, that what these theories say about the world is approximately correct. 

As introduced by \citep{feigl,horwich,van} and recently defined by \citep[p.10]{psillos}, semantic realism is associated with the realist’s desire to read physical theories literally, in the sense that the truth conditions of theoretical assertions are specified, not to be confused with epistemic realism, which consists in specifying their truth values (i.e., to assert whether or not the truth conditions obtain) \citep[p.10]{psillos}.\footnote{It must be stressed that without the semantic component, there cannot be an epistemic component. Indeed, although literal construals of theories are not the same as truth-value construals, it is necessary to have a literal interpretation of the theory just to specify the conditions that make its assertions true \citep{van}. } From the point of view of object-oriented realism, the semantic component stipulates that the theoretical terms of physical theories have the sole function of denoting putative objects, i.e, informing us what the world would be like according to these theories, \emph{as if they were correct}, whilst the epistemic component states, in addition to this semantic correspondence, that theoretical terms approximately refer to the actual objects that constitute the world. 

In the following section, we shall expand the definition of semantic realism by identifying additional semantic criteria against which our best physical theories (QM in particular) are to be evaluated. Since semantic realism has not been a point of special focus for philosophers of physics, we believe that they will find this way of expanding its definition both appropriate and fruitful. 
\section{Scrutinising semantic realism}\label{sec3}
According to our own reading of the above definition, we propose to identify two criteria that align with the semantic component of scientific realism: \emph{ontological clarity} and \emph{empirical adequacy}. 

On the one hand, ontological clarity means that the realist must specify what exactly the theory in question says (i.e., what its theoretical terms designate), in the sense that the realist must describe the scientific world depicted by the theory in terms of an ontology that has no considerable traces of semantic ambiguity. The association of this criterion with the semantic component of realism follows from conventional literature in the field, and in particular, from a proper reading of \citep{van}.\footnote{Again, we are citing the work of van Fraassen not because we are defenders of constructivist empiricism, but because we agree with the way he differentiates the semantic from the epistemic dimension of realism.} Saying, in van Fraassen’s words, that “a literal construal of a theory can elaborate by identifying what theoretical terms designate” is another way of saying that ontological clarity is, in part, a necessary condition for semantic realism. In particular, it is a condition that makes semantic realism different from epistemic realism because a literal construal is not related to “our epistemic attitudes towards theories, nor to the aim we pursue in constructing theories, but only to the correct understanding of what a theory says” \citep[11]{van}. Of course, as is the case for any kind of theoretical virtue, how to specify in any particular situation the precise conditions for ontological clarity to occur, is controversial. However, if we are intending to define ontological clarity from the perspective of physics, it can be conceived as providing a minimal characterisation of the ontological profile of the physical states of the theory, or what is frequently called the ‘ontic state’ of the theory. More specifically, we should clarify whether or not the mathematical objects defining the physical states denote something in the physical world, and if they do, what kind of entities they are supposed to be denoting.  

Empirical adequacy, on the other hand, means that the semantic realist also expects that the relevant empirical data, confined to the domain in which a physical theory can be effectively applied, can be explained in terms of the ontology in question. As was the case for ontological clarity, the latter definition of empirical adequacy is also prone to further concerns. More specifically, we should try to clarify the notions of ‘empirical data’ and ‘explanation’ involved in this definition. 

Firstly, note that the primary set of empirical data that scientists usually read from their complicated measuring devices before any kind of systematisation process takes place is a regularity occurring in the manifested world. Although this set of data represents information about the degrees of freedom of the system under analysis, this information is ultimately displayed in the three-dimensional macroscopic world in the form of some physical object and property, such as a ‘pointer’ heading in a certain direction. Thus, we can redefine empirical adequacy as a capability to explain the macroscopic manifestations of the relevant experimental data successfully predicted by the theory in terms of its posited (or elaborated) ontology. 

Secondly, the notion of explanation involved in the definition of empirical adequacy has to do with what is known as the ‘macro-object problem’,\footnote{This problem goes back to Locke’s `primary' and `secondary' properties and Eddington’s ‘two-table paradox’.} namely, the problem of accounting for how the qualities of macroscopic or observable objects (such as the colours of observable tables) are nothing but manifestations of the qualities of objects (such as electromagnetic waves interacting with tables made of atoms) that are ultimate and fundamental, out of which the actual structure of the world is constituted. To address this problem in the context of physics we have to fulfil two requirements, one technical, the other (metaphysically) explanatory. In the first place, we must show that any physical theory, if it deserves to be called a scientific theory, satisfies what has been called the ‘consistency constraint’ \citep{christian}. This technical requirement consists in ‘saving the phenomena’, namely, in identifying the theoretical basis required to make the predictions open to confirmation by a process of measurement. So, we should be capable of representing the underlying physical processes involved in typical measurements (and the outcomes produced by such measurements) in terms of the mathematical language of the theory. Considering the consistency constraint as an instrumental form of empirical adequacy (hereinafter, \emph{instrumental empirical adequacy}), its satisfaction is just a matter of modelling a mathematical entailment between a list of data and a set of theoretical principles and postulates. Since this requirement bears no resemblance to the nature and characterisation of the underlying ontology of the theory, it is necessary but not sufficient to satisfy the empirical adequacy criterion. On the other hand, we must satisfy an additional, different requirement: with perhaps the help of an ingenious metaphysical toolbox\footnote{We see that sometimes philosophers use metaphysical notions, such as supervenience, grounding or/and metaphysical dependence as part of this toolbox.}, we have to tell a consistent story of how the macroscopic world, such as tables and chairs, can be recovered from the fundamental (posited or elaborated) ontology of the theory \citep{ney0,ney1,christian}. This means that the macroscopic counterpart of the relevant experimental data successfully predicted by the theory must be cashed out in terms of its posited (or elaborated) ontology. This second requirement is, of course, related to the clarity criterion because we have to define a clear ontology before specifying how this ontology can comprehensively recover the observed phenomena. In this way, the clarity criterion is a necessary condition for the empirical adequacy criterion (understood in the broad metaphysical sense). 

We shall now proceed to briefly sketch the ‘logical’ structure of the quantum formalism which lies behind the measurement problem irrespective of the interpretation or philosophical position endorsed. After so doing, we shall investigate whether or not all the realist components are satisfied in the face of different interpretations of this formalism, with the aim of articulating the exact connection between scientific realism and the measurement problem. 
\section{The measurement process in quantum mechanics}\label{sec4}
First and foremost, we can think of the purely Hilbert space formalism of QM as an operational recipe solely designed for predictive purposes.\footnote{Sometimes, the Hilbert space formalism of QM is said to be purely logical or uninterpreted, but we believe that this way of characterising this formalism is misleading because although it has no correspondence with all the empirical consequences of the theory when a measurement process takes place (as we will show), it is already interpreted in terms of primitive (undefined) notions, such as  `physical states' and `observables', the structure of which we shall denote as ‘logical’.} As emphasised by \citep{maudlin2019}, this recipe does not necessarily commit itself to certain assumptions standardly associated with this theory. For example, the assumptions introduced as an interpretation of the purely Hilbert space formalism can be expressed in the form of truth-values assigned to the following claims:  
\begin{enumerate}
\item[(1)] The wave function of a system is complete (i.e., specifies all of the physical properties of a system). 
\item[(2)] The wave function always evolves in accordance with a linear, unitary dynamical equation (i.e., the Schrödinger equation). 
\item[(3)] The measurement of a physical property always has definite and single outcomes.\footnote{NB: do not confuse the claim that measurements have single outcomes with the claim that they have definite outcomes. As we shall see below, the first claim is related to the problem of outcomes, whilst the second one with the problem of interference.} 
\end{enumerate}
It is of course part of the discussion to be developed here whether or not some of these assumptions should be retained by this Hilbert space formalism of QM in order to obtain correct predictions. However, as long as these assumptions are not required to predict the outcomes that one would obtain in case a measurement is performed, they should be excluded from the predictive recipe. Considering this latter observation, we will sketch the possible ‘logical’ inferences that can be drawn (as regards the measurement process) from incorporating some of these additional assumptions to the purely Hilbert space formalism of QM without committing to any fully-fledged interpretation or philosophical position associated with this operational apparatus. In so doing, we shall revise the well-known account of the measurement process in QM arising from \citep{maudlin}. However, we shall only focus on its ‘logical’ construction as opposed to its interpretative consequences. Once the ‘logical’ structure of this account is revealed, in the next section we shall interpret it from the viewpoint of scientific realism as defined above. 

According to \citep{maudlin}, the ‘logical’ structure of the measurement process in QM can be presented in the form of the above three claims (1)-(3), all of which can be demonstrated to be mutually inconsistent. The proof of their mutual inconsistency will not be revised here as it is well known in the literature. The important point for us, however, is that this trilemma consists in a series of ‘logical’ inferences arising from incorporating additional assumptions to the operational apparatus of QM. As such, it cannot be considered problematic with the exception of evaluating the consistency of its ‘logical’ structure. This allows us to avoid the widespread mistake of interpreting the inconsistency between the above claims as representing the measurement problem without even providing the conceptual basis of what really counts as problematic. If there is something like a problem within this trilemma beyond the ‘logical’ consistency, we should pay attention to the possible interpretations associated with it. The introduction of an interpretation permits us to fix the truth-values of the above assertions together with a certain standard from which one evaluates the interpretation. It is precisely the identification of this standard that allows us to know in exact terms how exactly the measurement problem is associated with scientific realism. This will be the main task of the next section. 
\section{The problem of the measurement process in the face of scientific realism}\label{sec5}
Returning to our discussion, we shall now articulate the philosophical basis of the explanation of why and in what sense the incompatibility between completeness, unitary evolution, and definite-single values for measurement outcomes really constitutes a problem for QM. 

First and foremost, note that irrespective of Maudlin’s own interpretation, claim (3) can be construed \emph{empirically} as stating that the measurement of a physical property is always observed to have definite and single outcomes, or \emph{ontologically} as implying that, after the process of a measurement, there always exist objective definite and single outcomes for such a physical property. Considering that the observability of definite and single outcomes upon a measurement is a necessary (albeit not sufficient) condition for their existence, we shall interpret claim (3) ontologically. The reason for this particular interpretation (and the drawing of this ontological/empirical distinction) is mainly that the denial of the existence of definite and single outcomes upon measurement is necessary if incorporating NSQM into Maudlin’s framework, a set of interpretations that presumably solve the measurement problem and that we would like to critically discuss in this contribution. 

As we shall see next, we believe that the incompatibility of claims (1)-(3) is problematic if we are willing to interpret QM realistically on the basis of the criteria associated with the semantic component of realism (i.e., empirical adequacy and clarity with respect to the ontic state of the system in question). Let us now justify this claim by analysing the interpretative consequences of abandoning each one of the three claims (1)-(3), whilst retaining some of the others. For illustrative purposes we shall consider each of the relevant combinations of denying or accepting the above claims without following a predefined order of exposition. Although this analysis might seem a recitation of previous work, we believe that the association of Maudlin’s account of the measurement process with semantic realism has not been articulated. This association, as we shall see, shall permit us to characterise and address the corresponding problem from a broader, unambiguous standpoint.  
\subsection{The problem of instrumental empirical adequacy}
In the following lines, we shall argue that the measurement problem is associated with the lack of instrumental empirical adequacy that prevails in UQM, SQM, FRQM and NSQM. Whilst we shall concentrate a little bit more on demonstrating the lack of the instrumental form of empirical adequacy within FRQM and NSQM, it is important to address this task within UQM (which seems to be trivial) as we have to frame the well-known measurement process associated with QM in terms of our definition of instrumental empirical adequacy. 
\subsubsection{The instrumental empirical inadequacy of unitary quantum mechanics}
We shall demonstrate that if (3) is denied, whilst (1) and (2) are accepted, without embracing NSQM, we end up with a problem of instrumental empirical adequacy. Since by accepting (1) and (2) we are starting our analysis from a ‘logical’ structure which is based on a purely Hilbert space formulation, in addition to the completeness assumption, it would be wise to ask what exactly the mathematical formalism of vectors and unitary operators in Hilbert space has to do with the empirical adequacy of the theory, and therefore, with the particular physical situation of a typical measurement. To answer this question, let us characterise the notion of empirical adequacy in the particular context of QM.

Assuming that QM aims to describe the macroscopic world in terms of the way it describes the microscopic world (including the experimental devices in measurement situations), empirical adequacy states that the theory should have an operational technical procedure required to account for the underlying physical, macroscopic processes involved in typical measurements (i.e., instrumental empirical adequacy); and that the unobservable, microscopic ontology posited by the theory (or elaborated by the realist) should explain the empirical observations successfully predicted by it. These two basic criteria might be satisfied using different strategies. While the first one just would need a set of operational rules that dictate how the formalism of vectors and unitary operators in Hilbert space makes successful predictions about measurement outcomes (manifested in some way by the experimental device), the second one would require to figure out an appropriately grounded explanation (either physical or metaphysical) of how a set of definite and single outcomes of the observable properties associated with macroscopic objects (such as the direction of the pointer on the experimental device) correspond to the definite and single values of the properties associated with unobservable, microscopic objects (such as the spin of a particle, or the position associated with a system of particles, etc.). 

However, the perfect demonstration that the instrumental criterion of empirical adequacy cannot be satisfied by UQM corresponds to the famous example of the Schrödinger’s cat: UQM predicts that the quantum state of the whole system (represented by the wave function in a complete form), including the system under analysis and the experimental device, evolves unitarily according to the Schrödinger equation to an irreducible superposition of macroscopic states, where each non-realised term correlates, in a post-measurement situation, a definite and single value of the microscopic properties of the system under analysis (e.g., the atom decaying or not decaying) with the definite and single value of the corresponding macroscopic properties of the measurement device (e.g., the cat being dead or alive). This means that whatever the unobservable objects of the system, the predicted measured outcome recorded by the experimental device is undefined because it corresponds to a superposition of macroscopic states. Since every time a physical property is actually measured a definite and single outcome can be read from the experimental device in question, there is just a small step from this empirical fact to the conclusion that UQM is empirically inadequate in the instrumental form, in the sense that it does not have an operational procedure to explain the measurement process of systems in states of superposition. Therefore, since the semantic component of realism requires that UQM satisfies the empirical adequacy criterion in the instrumental form, and this criterion is not satisfied by it, then we are led to conclude that the acceptance of (1) and (2) and the abandonment of (3) is problematic by virtue of the fact that UQM cannot be interpreted realistically, at least in this semantic sense. 
\subsubsection{The instrumental empirical inadequacy of Frauchiger-Renner formulation}
Let us now address the most interesting situation. We shall demonstrate two conditionals at once: 
\begin{enumerate}
\item[] If claim (2) is denied, whilst (1) and (3) are retained, without embracing GRW (adding non-linear terms to the dynamics), we end up with a problem of instrumental empirical adequacy. \\
\item[] If claim (1) is denied, whilst (2) and (3) are retained, without embracing BQM (adding hidden variables to the physical state), we end up with a problem of instrumental empirical adequacy. 
\end{enumerate}
We shall define Frauchiger-Renner formulation (FRQM) as the theory that introduces into QM a distinction between two different points of view from which one describes the measurement process (hereafter \emph{Neumann's observer distinction}). As originally argued by \citep{neumann}, there is either the \emph{external} descriptive view, according to which the observer or the experimenter is treated in the same way as the system under analysis, evolving both by the Schrödinger (deterministic) law, or the observer’s descriptive \emph{internal} view, in which the system suddenly changes according to an indeterministic observer-dependent law (Born’s rule).\footnote{Note that this premise is sometimes expressed in terms of a distinction between internal observers located inside a laboratory that see definite and single outcomes and those external to it that see macroscopic superpositions. However, this way of making the aforementioned distinction is misguided in general: there can be internal observers seeing macroscopic superpositions in the same way that external observers are seeing definite and single outcomes. Granted this, the difference in notation between `internal' and `external observers' shall mean a distinction among observers who differ in relation to whether they see definite and single outcomes or a superposition, irrespective of whether they are inside or outside a laboratory.} This means that FRQM implicitly presupposes that there are external observers in FRQM that deny (1) and accept both (2) and (3), but also there are internal observers that deny (2) and accept both (1) and (3).\footnote{It is important to note that, according to our own reading of \citep{frauchiger}, Frauchiger and Renner assign truth-values to (1)-(3) claims without being aware that these claims are implicitly presupposed by them.} Granted this, we shall conclude that FRQM is empirically inadequate in the instrumental form. Let us justify this claim by considering internal observers first, and then proceed to introduce external observers (in addition to the internal ones). 

First and foremost, internal observers deny claim (2) (i.e., that the physical state always evolves unitarily according to the Schrödinger equation) by introducing an additional postulate into QM, called the \emph{projection postulate}. This collapse mechanism somehow makes the macroscopic superposition ``disappear'', such that one can effectively predict that the final state of the system after the measurement takes place is represented by the eigenvector corresponding to the measured outcome. At first sight, this strategy seems to be appropriate to satisfy the instrumental empirical adequacy criterion. Although this is true in typical cases, such as the ones with Schrödinger's cats, we believe that the first instrumental requirement of empirical adequacy cannot always be satisfied even with the introduction of the projection postulate. There is a concrete situation in which this scenario occurs: if we cannot say anything about the nature and behaviour of the physical system in question until it is observed then what has happened in the past ends up being beyond the descriptive scope of the theory. This situation, apart from being at odds with classical accounts of causality,\footnote{Some would justify this situation by assuming that the realisation of present facts causes facts in the past to be realised (an effect precedes its cause in time). This unfamiliar account of causality, called ‘retrocausality’, has been plainly discussed in \citep{backcausation}.} drastically reduces the empirical domain of the theory to the present state of affairs. Thus, FRQM is solely interpreted as a predictive recipe effectively applied in measurement situations, implying some temporal limitation in the way FRQM can be empirically adequate in the instrumental form. 
 
Secondly, recent thought experiments have demonstrated that the projection postulate, originally introduced by internal observers, is not always an appropriate tool for prediction because there are concrete physical systems with internal \emph{and} external observers upon which the application of this formal apparatus leads to inconsistent and incorrect predictions.\footnote{It is important to point out that this demonstration has been recently objected by \citep{okon2021}. Although he agrees that the projection postulate is not an appropriate tool for prediction for reasons concerned with its ambiguous nature, he argues that the conclusion arising from such thought experiments assume an incorrect assumption and this makes any prediction required by this conclusion unreliable. Since this assumption is not correct, according to him, there is no reason to conclude that the application of the projection postulate leads to inconsistent and incorrect predictions; on the contrary, it is because this postulate is so ambiguous that we mistakenly embrace incorrect assumptions and cannot even make predictions among these experiments. Although we think that Okon's objection should be discussed in depth in another paper, we shall assume that the projection postulate applied in the context of these thought experiments leads to inconsistent and incorrect predictions.} This shall lead to the conclusion that FRQM is empirically inadequate in the instrumental form. In order to clarify this point, we shall appeal to a well-known extension of the famous Wigner’s friend experiment, which is summarised as follows.

Recently introduced by \citep{frauchiger}, this controversial thought experiment (hereinafter the \emph{FR experiment}) is cast in the form of a no-go theorem (hereinafter the \emph{FR theorem}) that mainly states that no theory that is fully compliant with the predictions of QM can at once: (i) capture the universal validity of this theory (including the macroscopic level); (ii) assign a single outcome to each measurement; and (iii) demand consistency between different observers. The moral of this argument, according to \citep{frauchiger}, is that the mutual incompatibility of these assumptions puts forward the claim that FRQM “rejects a single-world description of physical reality” \citep[3]{frauchiger}. 

However, as critically discussed by numerous scholars \citep{sudbery,sudbery2,dustinhubert,tausk,mucinookon}, the FR theorem’s conclusion relies on a fourth silent, implicit assumption, which considerably reduces the scope of the theories ruled out by the FR theorem. Although this theorem is still a subject of controversy, there is a shared hypothesis accepted by (almost) the academic community that the alleged fourth silent assumption presupposed by the FR theorem corresponds to the aforementioned \emph{Neumann's observer distinction}. Since this assumption is obviously compatible with the projection postulate introduced by FRQM, this interpretation together with any other unknown interpretation that accepts (i)-(iii) together with the fourth silent assumption are ruled out by the FR theorem. In more precise terms, the point emphasised by these scholars is that this distinction presupposed by the FR theorem induces empirical differences between successive predictions obtained by internal and external observers. The reasoning is that if external observers would like to obtain further predictions after internal observers have performed their own measurements, and the former observers evolve the wave function according to Schrödinger’s equation and no term ever disappears, surely they must obtain different predictions than the latter observers would obtain applying the projection postulate (in which case some terms disappear).\footnote{Before discussing the FR theorem in detail, it is important to note that this conclusion relies on the specific form of the FR experimental arrangement. For example, it is well-known that in the \emph{double slit with monitoring experiment} interference effects are destroyed by the phenomena of entanglement and decoherence produced by the linear Schrödinger evolution, an empirical situation which is indistinguishable from the one obtained by applying the projection postulate.} Let us describe the FR theorem as it is discussed \citep{bub}, a version which will help us to better address the point at issue. 

Suppose there are two observers at a great distance from each other, Alice $A$ and Bob $B$, both of whom find themselves inside their respective laboratories, $A$  inside $L$ and $B$ inside $L'$. Additionally, there are two observers $X$ and $Y$ outside the respective laboratories, whom can perform measurements on them. For ease of notation, the irrelevant degrees of freedom of all the observers are not considered and the experimenters’ environment is excluded from the description, such as measurement devices and laboratory equipments. Before the experiment, $A$ prepares a coin $C$ in the initial state $\sqrt{1/3} \ket{h}_{A} + \sqrt{2/3} \ket{t}_{A}$.

Initially, at time $\tau=0$, $A$ measures $C$ in the $\{\ket{h}_{A},\ket{t}_{A}\}$ basis and records the result ``$h$'' or ``$t$''. At time $\tau=1$, she then prepares a qubit $Q$ as follows: if the result at $\tau=0$ came out to be $h$, $A$ prepares $Q$ in the state $\ket{0}_{B}$; if it rather came out to be $t$, she prepares it in the state $\frac{1}{\sqrt{2}}(\ket{0}_{B}+\ket{1}_{B})$. $A$ subsequently sends $Q$ to $B$ by a communication channel from $L$ to $L'$. At time $\tau=2$, $B$ measures $Q$ in the $\{\ket{0}_{B},\ket{1}_{B}\}$ basis and records the result ``$0$'' or ``$1$''. At subsequent times, there are two possible scenarios: 

The first scenario is such that at $\tau=3$ the observer $X$ measures $A$'s laboratory $L$ first in the basis:
\begin{eqnarray*}
\ket{f}_{L}=\frac{1}{\sqrt{2}}(\ket{h}_{A}+\ket{t}_{A})\\
\ket{o}_{L}=\frac{1}{\sqrt{2}}(\ket{h}_{A}-\ket{t}_{A})
\end{eqnarray*}
and records the result ``$o$'' or ``$f$'', whereas at $\tau=4$, the observer $Y$ measures $B$'s laboratory $L'$ in the basis:
\begin{eqnarray*}
\ket{f}_{L'}=\frac{1}{\sqrt{2}}(\ket{0}_{B}+\ket{1}_{B})\\
\ket{o}_{L'}=\frac{1}{\sqrt{2}}(\ket{0}_{B}-\ket{1}_{B})
\end{eqnarray*}
and records the result ``$o$'' or ``$f$''. On the contrary, the second scenario is such that the observer $Y$ measures first and precedes the measurement of the observer $X$. At the end of the experiment, the possible results of the measurements performed by each observer are compared. It turns out that a contradiction analogous to \citep{frauchiger} arises from this comparison. However, this contradiction is a conclusion derived from a series of premises that need to be explicated. We shall begin presenting a systematic formulation of the problem by enumerating both the explicit and hidden assumptions of the FR theorem in the following manner:
\begin{enumerate}
\item[] \emph{Premise (i)}: Universal validity of quantum mechanics. This means that QM describes both micro and macroscopic systems.\\
\item[] \emph{Premise (ii)}: Measurements have definite and single outcomes. \\
\item[] \emph{Premise (iii)}: Consistency between different observers. This means that if an observer $X$ finds that another observer $Y$ is certain about a measurement outcome, then $X$ must also be certain about such a measurement outcome. \\
\item[] \emph{Premise (iv)}: The Neumann’s observer distinction. This means that when a measurement is performed, some observers see definite and single outcomes, whereas others see macroscopic superpositions evolving unitarily.\\
\end{enumerate}
The contradiction arises in the following manner. Let us first consider (i) and (ii). The system of both laboratories $L$ and $L'$, composed of $A$, $B$, $C$ and $Q$, can be described by the following quantum state: 
\begin{equation}\label{eq1}
\ket{\Psi}=1/\sqrt{3} \left(\ket{h}_{A}\ket{0}_{B} + \ket{t}_{A}\ket{0}_{B} + \ket{t}_{A}\ket{1}_{B}\right)
\end{equation}
This expression is the state viewed from the perspective of any observer located outside the laboratories ($X$ or $Y$) without having described their own measurements. If we consider the measurements performed by $X$ in $L$ and $Y$ in $L'$, (\ref{eq1}) can be expressed as follows: 
\begin{eqnarray}
\ket{\Psi}&=&\frac{1}{\sqrt12} \ket{o}_{L}\ket{o}_{L'}-\frac{1}{\sqrt12} \ket{o}_{L}\ket{f}_{L'}\label{eq4}\\&+&\frac{1}{\sqrt12} \ket{f}_{L}\ket{o}_{L'}+\frac{1}{\sqrt12} \ket{f}_{L}\ket{f}_{L'}\nonumber\\
&=&\sqrt{\frac{2}{3}} \ket{f}_{L}\ket{0}_{B} + \frac{1}{\sqrt{3}} \ket{t}_{A}\ket{1}_{B}\label{eq2}\\
&=& \frac{1}{\sqrt{3}} \ket{h}_{A}\ket{0}_{B} + \sqrt{\frac{2}{3}} \ket{t}_{A}\ket{f}_{L'}\label{eq3}
\end{eqnarray}
Before any measurement is performed, the purely unitary evolution of the total composite system (i.e., $X$, $Y$, $A$, $B$, $C$, and $Q$) evolves to a superposition expressed by (\ref{eq4}), (\ref{eq2}), and (\ref{eq3}). Note that we can combine the results obtained by different observers at different times in the following way: 

From (\ref{eq2}) we can see that the relative state of $X$ and $Q$ with respect to the observer $X$ is $\ket{o}_{L} \ket{1}_{B}$ with the pair of possible values $\{o, 1\}$ (as a result of measuring $L$ and obtaining the outcome ``$o$''), whereas from (\ref{eq3}) we can see that the relative state of $C$ and $Y$ with respect to the observer $Y$ is $\ket{h}_{A} \ket{o}_{L'}$ with the pair of possible values $\{h, o\}$ (as a result of measuring $L'$ and obtaining the outcome ``$o$''). It follows that the pair of actual outcomes $\{h, 1\}$ are found only once the pair $\{o, o\}$ is obtained when $X$ and $Y$ jointly measure $L$ and $L'$, respectively. 

In this step we have used (iv), to assume that $X$ and $Y$ simultaneously obtained definite and single outcomes, so that we can predict future states of affairs conditioned by $XY$'s actual measurement outcome $\{o, o\}$; and also (iii), to be allowed to juxtapose $X$ and $Y$ to give us the perspective that consists of both observers $XY$. This latter assumption means that if $Y$ ($X$) finds that the $L$ ($L$') measurement outcome is ``$o$'' from the perspective of the $X$ ($Y$) observer, this outcome has also to be ``$o$''  as judged from the perspective of the $Y$ ($X$) observer, allowing us to observe the composed system $L L'$ from the $XY$ juxtaposed observer. Finally, only if we introduce preceding information of the state (\ref{eq1}), even after $A$, $B$, $X$, and $Y$ have measured $C$, $Q$, $L$, and $L'$, obtaining as a result definite and single outcomes (e.g., $h$ or $t$ for A; $1$ or $0$ for $B$; and ``$o$'' or ``$f$'' for $X$ and $Y$), we obtain a contradiction: the state (\ref{eq1}) shows that the probability of obtaining $\{h, 1\}$ is zero. 

Note that (iv) played an indispensable role in the previous argument because the final pair of outcomes $\{h, 1\}$ is a future prediction that was obtained only when the observer $XY$ used the collapsed state $\ket{o}_{L} \ket{o}_{L'}$ corresponding to what she found, not the full, unitary-evolving superposition (\ref{eq4}). At the same time, however, we considered the superposition (\ref{eq1}), corresponding to the unitary-evolving state of $A$, $B$, $C$, and $Q$ before any measurement takes place, even after the observer $XY$ performed a measurement and found $\{o, o\}$. 

Since FRQM assumes (iv) and the rest of the above premises, we may conclude that it is ruled out by the FR theorem. As regards the instrumental empirical adequacy of this interpretation, this conclusion brings important consequences. Note that the claim that FRQM leads to inconsistent predictions (as opposed to other interpretations, such as BQM) can be reinterpreted as stating that this collapse approach simply cannot make the correct quantum predictions in this experiment. Examples of these incorrect predictions are precisely those that were followed in the derivation of the contradiction, such as the possibility of predicting the final outcome $\{h, 1\}$ if the combined observer $XY$ measures the outcome $\{o, o\}$. In the framework of BQM, however, the outcome ``$o$'' of $Y$ is not only compatible with A’s outcome ``$h$'', as used in the FR theorem. Rather, the final outcome $\{t, 1\}$ is also possible. Let me elaborate more on this.

If there is an ontology constituted by space-time events in a non-relativistic context, such as configurations of Bohmian particles corresponding to $X$, $Y$, $A$, $B$, $C$, $Q$ and their respective settings and outcomes, the predictions obtained by each observer should, of course, respect the time order of those events. In the case in which observer $X$ measures first at time $\tau=3$ and $Y$ performs the second measurement at time $\tau=4$ (provided $A$’s and $B$’s measurements have been performed at time $\tau=0$ and $\tau=2$, respectively), $X$ will (locally) influence the subsystem composed of $A$ and $C$ in a way that $A$’s initial state and the corresponding result (either $h$ or $t$) could be changed. Since the state of $A$ and $B$ are correlated non-locally, if the result of $X$ turns out to be ``$o$'', it is a fact of the matter that $B$’s result must be $1$. This $1$ will remain invariant until the second measurement of observer $Y$ is performed. This second measurement could (locally) influence and change $B$’s state and its corresponding result (either $1$ or $0$). Now, in this story it is not true that the result ``$o$'' of $Y$ is only compatible with $A$’s result $h$, as it would be in the case in which observer $Y$ measures first and $X$ performs the second measurement. In sum, in the case in which $Y$ performs her measurement first, $Y$’s outcome is independent of $X$’s setting and outcome; in the case in which $X$ performs his measurement first, however, $Y$’s outcome depends on $X$’s setting and outcome.

Therefore, since BQM is not ruled out by the FR theorem and does not lead to inconsistencies, we may reinterpret the inconsistencies arising from FRQM as the failure of this approach to make some correct quantum predictions. This makes FRQM theory empirically inadequate in the instrumental form. That is, FRQM is ruled out by the FR theorem by virtue of the fact that it cannot account for FR experiment's predictions correctly obtained by, for example, BQM. 

Considering this line of reasoning, one might wonder whether the hidden assumption (iv) is the ultimate reason FRQM cannot make the correct predictions in the FR experiment. Following a similar argument to \citep{okon2021} (although not the same conclusion), we believe that there is a deeper reason that lies behind (iv) that is responsible for the instrumental empirically inadequacy of FRQM in the context of the FR experiment. This underlying reason is associated with the fact that $X$ or $Y$ does not possess the resources to predict the complete outcomes obtained by $A$ and $B$ after the outcomes obtained by the juxtaposed observer $XY$ are determined. Since the state (\ref{eq1}), as defined from the point of view of $X$ or $Y$ (without having described their own measurements), evolves according to the linear Schrödinger evolution, and all observers (i.e., $X$, $Y$, $A$ and $B$) obtain definite and single outcomes, Maudlin's trilemma tell us that FRQM is incomplete. In other words, the state (\ref{eq1}) cannot specify all the physical properties of $A$ and $B$. It is expected that the lack of specification by the wave function of the actual physical properties associated with these subsystems induces the alleged incorrect predictions and explains why BQM, which is a complete theory, makes the correct predictions. 

This result reflects the fact that in the FR experiment there are not only internal observers that deny (2) and accept (1) and (2), but also there are external observers that deny (1) and accept (2) and (3). In other words, FRQM is empirically inadequate in the instrumental form precisely because there are experimental situations in which it abandons the completeness assumption.
\subsubsection{The instrumental empirical inadequacy of non-collapse interpretations}\label{noncollapse}
We shall now demonstrate that if (3) is denied, whilst (1) and (2) are retained, we end up with the same problem of instrumental empirical adequacy (as corroborated by the previous discussion) unless we explain the empirical fact that measurement outcomes \emph{seem} to be definite and single-valued even if macroscopic superpositions exist and never really collapse, an explanation which will be shown to be unsuccessful. Considering that claim (3) presupposes the existence of (as opposed to the observation of apparent) definite and single outcomes, its denial leads therefore to the adoption of NSQM.\footnote{NSQM will be defined here simply as those interpretations which deny the existence of definite and single outcomes but retain (1) and (2). In this way, we shall include neither BQM nor the Hamiltonian modal interpretation as part of what we call NSQM because these interpretations accept the existence of definite and single outcomes through the introduction of a preferred set of commuting observables (i.e., the position and the energy, respectively).} 

One might argue that denying claim (3), whilst retaining (1) and (2), is one exception to the conclusion that QM is empirically inadequate in the instrumental form. This exceptional case would derive from simply denying the existence of definite and single outcomes and accepting the existence of macroscopic superpositions, and therefore concluding that the theory is empirically adequate in the instrumental form because it naturally accounts for actual superpositions. However, the instrumental empirical adequacy of a theory cannot be derived from simply assuming that it accounts for something that exists but is impossible to detect in everyday experience, such as macroscopic superpositions. Let us appeal to an example which corroborates this claim. Based on numerous recent ‘Schrödinger cat states’ experiments, some philosophers have proffered NSQM according to which macroscopic (in addition to microscopic) superpositions are objective physical states which never really collapse. Examples of NSQM are the many-worlds interpretation, relational quantum mechanics, the modal interpretation and the consistent histories approach. However, although the way to account for the reality of these macroscopic superpositions varies across the board, the common feature among these interpretations is that they believe that measured outcomes are not actually definite and single-valued even if we observe that they seem to be definite and single-valued in everyday experience. That is, the price to be paid for denying claim (3) whilst retaining (1) and (2) is that there should be a way to explain the observable fact that macroscopic superpositions are impossible to detect in everyday experience and, when the properties of a physical system are measured, measurement outcomes \emph{seem} to have definite and single values. This example demonstrates that even when the existence of macroscopic superpositions is assumed, whilst that of definite and single outcomes is denied, the challenge of instrumental empirical adequacy still prevails. Unfortunately, NSQM are not empirically adequate in the instrumental form, as we shall demonstrate next. We believe that NSQM are empirically inadequate in the instrumental form for at least two possible reasons. Whilst we accept that the first reason might be argued to be ungrounded for certain reasonable intuitions (to be described below), we believe that the second one poses a serious challenge to the instrumental empirical adequacy of these interpretations. 

Firstly, one might argue that the FR theorem can also be applied to NSQM based on the possibility that these interpretations comply with all (i)-(iv) assumptions, and introduce internal observers that deny (3) and accept (1) and (2) in addition to external observers that deny (3) and (1) and accept (2). In a similar way to FRQM, this would lead to the conclusion that NSQM are ruled out by the FR theorem, and therefore are empirically inadequate in the instrumental form by virtue of the fact that they yield incorrect predictions. To arrive at this conclusion, one would need to focus on the way assumption (iv) is interpreted. Note that the aforementioned Neumann’s observer distinction between different observers that lies behind assumption (iv) might be interpreted epistemically (as opposed to ontologically). Since macroscopic superpositions exist and never really collapse in NSQM, the fact that definite and single outcomes are observed might be interpreted as an apparent \emph{epistemic} manifestation, not necessarily associated with \emph{actual}, definite and single outcomes. This would mean that in the case of NSQM, this hidden assumption is not precisely aimed at reconciling unitary evolution with objective (actual) definite and single outcomes, such as our own reading of \citep{maudlin}’s analysis in denying completeness; on the contrary, it interprets both unitary evolution and the epistemic (empirical) fact that definite and single outcomes are seemingly observed as two different, non-reducible points of view. Therefore, the fourth silent assumption of the FR theorem would not only be compatible with particular, idiosyncratic interpretations of QM that believe in the existence of definite and single outcomes, such as FRQM, but also with some NSQM that explain in some way or other the appearance of (apparent) definite and single outcomes, such as relational quantum mechanics, the consistent histories approach and some modal interpretations. 

Unfortunately, this conclusion might be argued to be ungrounded after a careful analysis. In fact, one might reply that the compatibility between NSQM and assumption (iv) is not sufficient to derive the conclusion that these interpretations are ruled out by the NSQM theorem. Rather, they should be compatible with the rest of the assumptions (i)-(iii). But we are currently uncertain of whether or not these interpretations accept the other assumptions. For example, based on an intuitive reading of some of these interpretations, one might argue that relational quantum mechanics denies (iii), whilst the consistent histories approach denies (ii), and therefore block the conclusion that they are ruled out by the FR theorem.\footnote{One concrete idea in this direction is perspectivalism endorsed by \citep{dieks}, which mainly states that NSQM are interpretations that deny (iii) by virtue of the fact that they permit the possibility of ascribing to the same physical system more than one objective state.} However, as things currently stand, we cannot undoubtedly conclude, with the exception of FRQM, that this is actually the case.

Secondly, even if we were to block the previous conclusion in solid ground, we would inevitably came to realise that NSQM are empirically inadequate in the instrumental form because they have not succeeded in solving at least one problem, the issue of selecting a single outcome upon the measurement of a physical property (hereinafter the \emph{problem of outcomes}).\footnote{There is another issue associated with instrumental empirical adequacy, that of explaining the need to describe the theory in terms of a privileged basis, normally associated with the position. However, since our aim is to provide arguments \emph{contra} the claim that the theory is empirically adequate in the instrumental form, we do not need to discuss the problem of the basis if we demonstrate that the problem of outcomes, which is a necessary condition for this form of empirical adequacy, cannot be solved.} 

As regards the instrumental criterion of empirical adequacy, the principal challenge is to provide an empirical interpretation of the quantum formalism without collapses associating mathematical representations (e.g., vectors and operators) with the empirical data. To have this interpretation, however, the purely unitary evolution of this formalism should not only be able to macroscopically distinguish the mathematical counterparts of the possible independent outcomes associated with the physical properties of the system (hereinafter the \emph{problem of interference}), but also to pick up a single outcome that is actually measured (i.e., the problem of outcomes, as defined above). Whereas the first problem is addressed by providing a recipe to reduce the overlap between the effective wave function packets of the macroscopic superposition (as long as they are allowed to have approximately disjoint support), the second is overcome by providing a selective method to pick up the respective wave packet associated with the measured outcome. 

One implicit and essential method to which some NSQM usually appeal is the decoherence mechanism.\footnote{Although the decoherence mechanism is regularly involved in these interpretations, as is argued in \citep{baq}, it is not necessarily involved, as we shall see below.} From the point of view of decoherence, the process of measurement consists in a (purely unitary) quantum-mechanical interaction between the system and a great number of environmental degrees of freedom (including those of the observer and the measurement apparatus) that come into play, such that the interference of macroscopic states exists but cannot be observed. Formally speaking, the decoherence mechanism treats, for all practical purposes, the improper or reduced states of the superposition \emph{as if they were} proper mixed states. This allows the identification of a set of distinguishable wave packets with approximately disjoint support (enough to be macroscopically relevant) associated with the possible values of the physical quantities to be measured (together with their probability of occurrence). 

However, it is important to remark that the decoherence mechanism does not (and does not aim to) select a single outcome during the process of measurement. Since decoherence only intends to solve the problem of interference (as opposed to the problem of outcomes), the NSQM that appeal to decoherence need to account for their instrumental empirical adequacy by other means. This does not mean, however, that if decoherence were reliable, it would be useless as regards the instrumental empirical adequacy of the theory. As is correctly pointed out by \citep{schloss}, there is no way of solving the problem of outcomes without accounting for the non-observability of interference. This means that adherents of decoherence agree that there is a special mechanism described by this framework that is necessary (but not sufficient) for the observer to see definite and single outcomes.\footnote{Although not generally accepted, there have been serious objections to considering decoherence a reliable strategy not only to select a single outcome (something which is generally accepted as noted above) but even to explain the non-observability of interference. One of these objections was recently proposed by \citep{okonsudarsky}.} 

Under these circumstances, proponents of NSQM have figured out other strategies that might solve the problem of interference together with the problem of outcomes without appealing to the decoherence mechanism. For example, let us consider relational quantum mechanics. As discussed in \citep{Rovelli1996}, the apparent breakdown of the unitary evolution of the physical state results, according to this interpretation, from the fact that the observer is unable to provide a complete description of its interaction with the system. This relational approach is based on a theorem due to \citep{Breuer1995}, which states that if a measuring apparatus is part of a physical system, the state of the apparatus cannot fully encode the state of the whole system (including the apparatus itself). Consequently, we cannot infer the state of the whole system by observing the state of the apparatus. Although we shall not inquire into the details of relational quantum mechanics, it is important to note that this approach has fallen prey to various objections, some of which point towards the lack of ontological clarity and empirical adequacy of this theory.\footnote{One example is due to \citep{mucino2021}. However, Rovelli’s response to this objection \citep{Rovelli2021} has been criticised because he does not offer a solid argument that addresses the point at issue.} In replying to these critiques, one might suggest introducing the decoherence mechanism to solve this problem of clarity and empirical adequacy (as seems to be suggested in \citep{DiBiagio2020}). However, if we reconsider this strategy, we fall back into the above objections. It seems that NSQM cannot get rid of the decoherence mechanism until they show that it is irrelevant to them without facing serious problems, an approach which, to the best of our knowledge, has not been entirely successful.\footnote{Another NSQM that does not appeal to the decoherence mechanism but has not escaped serious objections is the modal interpretation according to which definite properties depend on the quantum state (through what is called the bi-orthogonal (Schmidt) decomposition) and change over time \citep{vermasdieks}.}

Omitting further details, the most important point to learn from this discussion is the fact that, without providing a reliable answer to the problem of outcomes, the instrumental empirical adequacy of NSQM, which is a necessary condition for scientific realism, is compromised. Specifically, these theories, as they are standardly formulated without any collapse mechanism, do not have an operational technical procedure required to account for definite and single outcomes. 
\subsubsection{The instrumental empirical inadequacy of epistemic/statistical interpretations}
Finally, we shall see that if claim (1) is denied, whereas (2) and (3) are retained, without adding hidden variables to the physical state, then we end up with a problem of instrumental empirical adequacy. As we know from \citep{maudlin}, denying claim (1) without introducing hidden variables leads to epistemic/statistical interpretations.  

According to many scholars, there is a shared intuition that epistemic/statistical interpretations are irrelevant for a quantum realist reading simply because they are only interested in calculating predictions or the lack of knowledge of the physical system as opposed to describing objective states of affairs through the lens of QM. Our opinion is that this intuition is true for statistical but not for epistemic interpretations. The problem with the latter interpretations is that if we are aiming at calculating the lack of knowledge of certain states of affairs, we need to have at least a predetermined standard in physical reality with respect to which we are making such a calculation. Indeed, epistemic interpretations make a silent and unjustified commitment to the existence of some unknown, predetermined aspect of the physical world, whilst they explicitly deny the possibility that QM may describe that aspect of the world objectively. 

Let us illustrate this critique with an example provided by \citep{maudlin} without inquiring into its details. If the state of the system composed of the spin of an electron and an x-spin measuring device evolves to a macroscopic superposition (as dictated by QM), epistemic interpretations will interpret this superposition as if it \emph{really} were a mixed state, meaning that the whole system is in either one or the other term of the superposition, but we are uncertain about which is the actual realised term. However, in the absence of any justification for regarding a pure state as a mixture, the macroscopic superposition alone cannot be interpreted in epistemic terms. The reason is that we are incapable of being ignorant of which individual term is realised if we cannot differentiate and single out any term from the superposition to begin with. In this way, the identification of terms in this macroscopic superposition would be equally regarded as introducing predetermined values for the spin of the electron that are not included in (or described by) QM. This seems not irrelevant but a problem for at least the semantic component of scientific realism. Let us recall that scientific realism in general and semantic realism in particular presuppose the possibility of providing an ontologically clear and empirically adequate description of the physical world \emph{through the lens of QM}. However, although epistemic interpretations are committed to the existence of the aspect of the world of which we are ignorant ---which means that they accept the mind-independent condition of scientific realism--- they assume that no description of this aspect of the world can be provided by QM. This leads to the conclusion that these interpretations are empirically inadequate in the instrumental form. Let us elaborate more on this. 

The first obvious consequence of endorsing the epistemic interpretation is that, following the previous example, it does not provide a solution to the aforementioned problem of outcomes ---a sufficient condition for the instrumental empirical adequacy of QM--- as this interpretation explicitly denies the possibility of knowing with certainty which is the realised, single outcome of the electron spin. Furthermore, the same example shows that epistemic interpretations do not provide a solution to the aforementioned problem of interference ---a necessary condition for the instrumental empirical adequacy of QM--- because they presuppose that the physical state of the electron spin and the measuring device represent an array of possible definite outcomes, without providing any justification for this presupposition based on QM. Therefore, the problem of instrumental empirical adequacy remains unsolved. 
\subsection{The problem of ontological clarity}
As we shall see, the measurement problem is also related to the lack of ontological clarity regarding the interpretation of the theory. The reason for this relation can be traced back to the empirical adequacy criterion broadly understood. Recall that ontological clarity is a necessary condition for empirical adequacy. The latter not only involves an instrumental criterion (the consistency constraint), but also the (metaphysically) explanatory criterion, which depends on the clear specification of the ontic state of the theory. Thus, regardless of the satisfaction of the instrumental criterion, the fact that a theory is ontologically unclear implies that it is at odds with the explanatory criterion, and thus with the empirical adequacy criterion. To illustrate this observation, let us investigate each of the interpretations already mentioned.
\subsubsection{The lack of ontological clarity of unitary quantum mechanics}
We shall demonstrate that if claim (3) is denied but (1) and (2) are retained, without embracing NSQM, we end up with a problem of ontological clarity. This shall lead to the conclusion that UQM is ontologically unclear.  

In addition to the deterministic evolution, the root of the instrumental empirical adequacy problem in UQM lies on the basic realist assumption of this theory that the physical state of the system is complete and represents in some way or other all of the objective physical properties of the system. In this way, it is assumed that the macroscopic superposition of states that arises in measurement situations is a physical state in its own right, in the sense that it represents the quantum world as it, in fact, is. However, note that QM does not specify with precision the ontic state of the theory. That the wave function specifies all the physical properties of a system does not imply that it exists as an object in the world. Rather, it means that there are physical properties of unobservable entities (not explicitly specified) that have a factual correspondence with the wave function. Thus, as things stand, UQM is unclear in the ontological sense. 
\subsubsection{The lack of ontological clarity of collapse and non-collapse interpretations}
We shall demonstrate three conditionals at once: 
\begin{enumerate}
\item[] If claim (1) is denied, whilst (2) and (3) are retained, without embracing BQM (adding hidden variables to the physical state), we end up with a problem of ontological clarity. \\
\item[] If claim (2) is denied, whilst (1) and (3) are retained, without embracing GRW (adding non-linear terms to the dynamics), we end up with a problem of ontological clarity.\\
\item[] If claim (3) is denied, whilst (1) and (2) are retained, we end up with a problem of ontological clarity.
\end{enumerate}
Since the first two conditionals are embraced by FRQM; the second conditional is embraced by SQM; and the third one by NSQM, we shall conclude that these three theories and interpretations are ontologically unclear.

As regards SQM and FRQM, the introduction of the projection postulate means that what was initially a deterministic and unitary evolution governed by the Schr\"odinger equation turns out to be an indeterministic jump that has to be introduced by hand from the outside. The lack both of naturalness and a clear and comprehensive explanation for this collapse mechanism is the reason why many philosophers think of the measurement problem in SQM and FRQM as a lack of objectivity with respect to our theoretical dealings with the quantum phenomena. According to this scenario, both theories would be at odds with our objectivity standards because the nature and behaviour of what is measured is observer-dependent. There would not be any kind of explanation of what happens when the observer does not have a human brain capable of rationalising like a PhD student, or when the observer is not present as a conscious witness at a certain place and time \citep{bellhidro}. In the particular case of FRQM, this problem has been characterised as a conflict (or lack of rules) in establishing the correct interpretation among the two different points of view arising from the aforementioned Neumann’s observer distinction. Furthermore, the epistemological or ontological interpretation of this conflict between two different points of view has been taken as the basis for further reasons why this theory must either objectively describe the state of consciousness of the observer \citep{london}.

What about NSQM? Although these interpretations do not introduce the projection postulate and the system in NSQM evolves according to a purely unitary evolution, the aforementioned Neumann’s observer distinction between different observers is also introduced by means of certain external mechanism (such as decoherence) that makes possible the observability of (epistemic) definite and single outcomes. This means that, apart from the obscure character of this or other proposed non-collapse mechanisms, NSQM preserve the same problems of SQM and FRQM as regards objectivity. 

Consequently, what lies behind these alternative views or conclusions is the fact that the nature and the dynamics of the physical state of the system alone cannot account for (objective or epistemic) definite and single measurement outcomes and because of that, SQM, FRQM and NSQM must appeal to ambiguous mechanisms whose operation does not depend on the objective constituents, principles and laws of the theory but on the idiosyncrasy of the observer. With the projection postulate or any non-collapse mechanism introduced, however, these theories remain ontologically unclear in the sense that there is no pre-conceived objective ontology that can exhaustively explain the observed phenomena and the measurement process itself as any objective physical interaction between the constituents of the system and the experimental device in question. Rather, the constitutive nature and behaviour of what is measured directly depends on the experimental context and cannot be conceived before the interaction takes place. In this way, the projection postulate or any non-collapse mechanism involve satisfaction of neither the instrumental criterion (as demonstrated above) nor the explanatory criterion, both of which are necessary to make the theory empirically adequate. Therefore, since the semantic component of realism requires that SQM, FRQM and NSQM satisfy the clarity criterion, and this criterion is not satisfied by them, then we are led to conclude that they cannot be interpreted realistically, at least in this semantic sense.  

Furthermore, the fact that FRQM is ontologically unclear not only implies that this theory is at odds with our preferred definition of scientific realism, but also that the first instrumental requirement of empirical adequacy cannot always be satisfied even with the introduction of the projection postulate. Indeed, the FR theorem described above may be regarded as an illustration of the claim that instrumental empirical adequacy fails as a consequence of the ambiguity of the projection postulate (or any mechanism that explains the observability of definite and single outcomes). The reason is that the acceptance of Neumann’s observer distinction by FRQM implements a profound ontological ambiguity upon the description of the quantum phenomena. Thus, in order to avoid incorrect predictions we have to know how to apply the formalism of FRQM in a given physical situation. We suggest that it is through the specification of a clear ontology and the acquisition of knowledge of what exactly is happening in the physical world when a measurement is performed that we are able to apply this formalism without inconsistencies. As is the case for the second metaphysical criterion of empirical adequacy, the FR theorem also suggests that the first instrumental criterion cannot be satisfied without the ontological clarity criterion. Empirical adequacy and ontological clarity are deeply correlated semantic realist criteria. 
\subsubsection{The lack of ontological clarity of epistemic/statistical interpretations}
Finally, we shall see that if claim (1) is denied, whereas (2) and (3) are retained, without embracing BQM (adding hidden variables to the physical state), then we end up with a problem of ontological clarity associated with epistemic/statistical interpretations.  

Statistical interpretations are by construction ontologically unclear by virtue of the fact that they are only interested in calculating predictions of the physical system. As for epistemic interpretations, let us recall the famous PBR theorem. According to certain (debatable) assumptions, this theorem shows that models of QM in which pure quantum states are interpreted epistemically (i.e., represent probabilistic or incomplete states of knowledge about physical reality) cannot reproduce quantum predictions. As a result, pure quantum states must be construed ontologically in the sense that they describe objective states of affairs. As is well known, this conclusion is derived by considering that the quantum state can be interpreted ontologically if “every complete physical state or ontic state in the theory is consistent with only one pure quantum state”; or epistemically ``if there exist ontic states that are consistent with more than one pure quantum state" \citep{PBR}. This distinction lends support to the idea that epistemic quantum states cannot describe ontic states. Furthermore, if we were to interpret the quantum state epistemically, we would be incapable of providing a minimal characterisation of the ontological profile of the ontic state of the theory. Since this minimal characterisation is a prerequisite for the ontological clarity of QM, the epistemic interpretation, according to this distinction, neglects the possibility of QM being ontologically clear and, therefore, does not satisfy the semantic component of scientific realism. We can deduce from this observation that if the premises of the PBR theorem are true, its conclusion reinforces the objection that epistemic interpretations pose against ontological clarity, and provides necessary (albeit not sufficient) conditions to be a semantic realist with respect to QM. 
\section{Concluding remarks}\label{sec6}
Based on a well-known characterisation of scientific realism, we conclude that the description of the processes involved in typical measurements framed in terms of UQM, SQM, FRQM, NSQM and epistemic and statistical interpretations is problematic from the standpoint of what we have defined as semantic realism. In particular, we have argued that this description is problematic by virtue of the fact that there is no specification of what exactly these interpretations say in terms of their ontologies (i.e., they are ontologically unclear), and that the relevant empirical data, confined to the domain in which they can be effectively applied, cannot be explained in terms of these ontologies and cannot be entailed by their corresponding formulations (i.e., they are empirically inadequate). 


\begin{thebibliography}{00}
\bibitem[Bacciagaluppi(2016)]{baq}
Bacciagaluppi, G. (2016). The role of decoherence in quantum mechanics. In E. N. Zalta (Ed.), \emph{The Stanford Encyclopedia of Philosophy, Fall 2016 Edition}. Metaphysics Research Lab, Stanford University. URL=$<$https://plato.stanford.edu/archives/fall2016/entries/qm-decoherence/$>$.
\bibitem[Bell(1971)]{bellhidro}
Bell, J. S. (1971). Against measurement. In Aspect, A. (Ed.) \emph{Speakable and Unspeakable in Quantum Mechanics} (pp. 213-231). Cambridge: Cambridge University Press.
\bibitem[Bohm(1952)]{bohm}
Bohm, D. (1952). A Suggested interpretation of the quantum theory in terms of ‘hidden’ variables. I and II. \emph{Physical Review}, \emph{85}(2), 166-193. 
\bibitem[Breuer(1995)]{Breuer1995}
Breuer, T. (1995). The impossibility of accurate state self-measurements. \emph{Philosophy of Science}. \emph{62}, 197-214. 
\bibitem[Bub(2017)]{bub}
Bub, J. (2017). Why Bohr was (mostly) right. \emph{arXiv:1711.01604v1[quant-ph]}.
\bibitem[Chakravartty(2017)]{real}
Chakravartty, A. (2017). Scientific Realism. \emph{The Stanford Encyclopedia of Philosophy} (Summer 2017 Edition), Edward N. Zalta (ed.), URL=$<$https://plato.stanford.edu/archives/sum2017/entries/scientific-realism/$>$. 
\bibitem[Biagio and Rovelli(2020)]{DiBiagio2020}
Di Biagio, A., \& Rovelli, C. (2020). Stable facts, relative facts. ArXiv:2006.15543. 
\bibitem[Dieks(2019)]{dieks}
Dieks, D. (2019). Quantum mechanics and perspectivalism. In Lombardi, O., Fortin, S., López, C., \& Holik, F. (Eds.) \emph{Quantum Worlds: Perspectives on the Ontology of Quantum Mechanics} (pp. 51-70). Cambridge: Cambridge University Press.
\bibitem[D\"{u}rr et al.(1992)]{durr}
Dürr, D., Goldstein, S., \& Zanghì, N. (1992). Quantum equilibrium and the origin of absolute uncertainty. \emph{Journal of Statistical Physics}, \emph{67}(5), 843-907.
\bibitem[Faye(2018)]{backcausation}
Faye, J. (2018). Backward Causation. \emph{The Stanford Encyclopedia of Philosophy} (Summer 2018 Edition), Edward N. Zalta (ed.), URL = $<$https://plato.stanford.edu/archives/sum2018/entries/causation-backwards/$>$. 
\bibitem[Feigl(1950)]{feigl}
Feigl, H. (1950). Existential hypotheses: realistic versus phenomenalistic interpretations. \emph{Philosophy of Science}, \emph{17}, 35-62.
\bibitem[Frauchiger and Renner(2018)]{frauchiger}
Frauchiger, D. \& Renner, R. (2018). Quantum theory cannot consistently describe the use of itself. \emph{Nature Communications}, \emph{9}(3711). 
\bibitem[Ghirardi(2016)]{girardi}
Ghirardi, G. (2016). Collapse Theories, \emph{The Stanford Encyclopedia of Philosophy} (Spring 2016 Edition), Edward N. Zalta (Ed.), URL = $<$https://plato.stanford.edu/archives/spr2016/entries/qm-collapse/$>$. 
\bibitem[Holland(1993)]{holland}
Holland, P. (1993). \emph{The Quantum Theory of Motion: An Account of the de Broglie-Bohm Causal Interpretation of Quantum Mechanics}. Cambridge: Cambridge University Press.
\bibitem[Horwich(1982)]{horwich}
Horwich, P. (1982). Three forms of realism. \emph{Synthese}, \emph{51}(2), 181-201.
\bibitem[Lam and W\"uthrich(2018)]{christian}
Lam, V., \& W\"uthrich, C. (2018). Spacetime is as spacetime does. \emph{Studies in History and Philosophy of Modern Physics}, \emph{64}, 39-51.
\bibitem[Lazarovici and Hubert (2019)]{dustinhubert}
Lazarovici, D., \& Hubert, M. (2019). How quantum mechanics can consistently describe the use of itself. \emph{Scientific Reports}, \emph{9}(470).
\bibitem[London and Bauer(1939)]{london}
London, F., \& Bauer, E. (1939). The Theory of Observation in Quantum Mechanics. In Wheeler, J. \& W. Zurek, W. (Eds.) \emph{Quantum Theory and Measurement} (pp. 217-259). New Jersey: Princeton University Press. 
\bibitem[Maudlin(1995)]{maudlin}
Maudlin, T. (1995). Three measurement problems, \emph{Topoi},14, 7-15.
\bibitem[Maudlin(2019)]{maudlin2019}
Maudlin, T. (2019). \emph{Philosophy of physics: Quantum theory}. Princeton: Princeton University Press.
\bibitem[Muciño et al.(2021)]{mucino2021}
Muciño, R., \& Okon, E., \& Sudarsky, D. (2021). Assessing relational quantum mechanics. arXiv:2105.13338 [quant-ph]. 
\bibitem[Muciño and Okon(2020)]{mucinookon}
Muciño, R., \& Okon, E. (2020). Wigner's convoluted friends. \emph{Studies in History and Philosophy of Modern Physics}. \emph{72}, 87-90.
\bibitem[Ney(2015a)]{ney0}
Ney, A. (2015). Fundamental physical ontologies and the constraint of empirical coherence: a defense of wave function realism. \emph{Synthese}, \emph{192}(10), 3105-3124.
\bibitem[Ney(2015b)]{ney1}
Ney, A. (2017). Finding the world in the wave function: some strategies for solving the macro-object problem. \emph{Synthese}. https://doi.org/10.1007/s11229-017-1349-4.
\bibitem[Okon(2021)]{okon2021}
Okon, E. (2021). On the objectivity of measurement outcomes. \emph{The British Journal for the Philosophy
of Science} (forthcoming).
\bibitem[Okon and Sudarsky(2016)]{okonsudarsky}
Okon, E., \& Sudarsky, D. (2016). Less decoherence and more coherence in quantum gravity, inflationary cosmology, and elsewhere. \emph{Foundations of Physics}, 46(7), 852-879.
\bibitem[Psillos(1999)]{psillos}
Psillos, S. (1999). \emph{Scientific Realism: How Science Tracks Truth}. London: Routledge.
\bibitem[Pusey et al.(2012)]{PBR}
Pusey, M., \& Barrett, J., \& Rudolph, T. (2012). On the reality of the quantum state. \emph{Nature Physics}. \emph{8}, 475-478. 
\bibitem[Rovelli(1996)]{Rovelli1996}
Rovelli, C. (1996). Relational Quantum Mechanics. \emph{International Journal of Theoretical Physics}. \emph{35}(8), 1637-1678. 
\bibitem[Rovelli(2021)]{Rovelli2021}
Rovelli C. (2021). A response to the Muciño-Okon-Sudarsky's assessment of relational quantum mechanics. ArXiv:2106.03205 [quant-ph].
\bibitem[Schlosshauer(2008)]{schloss}
Schlosshauer, M. (2008). \emph{Decoherence and the Quantum-to-Classical Transition}. Heidelberg: Springer.
\bibitem[Sudbery(2017)]{sudbery}
Sudbery, A. (2017). Single-World Theory of the Extended Wigner’s Friend Experiment. \emph{Foundations of Physics}. \emph{47}, 658-669. 
\bibitem[Sudbery(2019)]{sudbery2}
Sudbery, A. (2019). The hidden assumptions of Frauchiger and Renner. \emph{International Journal of Quantum Foundations}. \emph{5}(3), 98-109.
\bibitem[Teller(1994)]{tausk}
Tausk, D. (2019). A brief introduction to the foundations of quantum theory and an analysis of the Frauchiger-Renner paradox. arXiv:1812.11140v2. 
\bibitem[van Fraassen(1980)]{van}
van Fraassen, B. (1980). \emph{The Scientific Image}. Oxford: Clarendon Press Oxford.
\bibitem[Vermaas and Dieks(1995)]{vermasdieks}
Vermaas, P.,  \& Dieks, D. (1995). The modal interpretation of quantum mechanics and its generalization to density operators. \emph{Foundations of Physics}. \emph{25}, 145-158.
\bibitem[von Neumann(1932)]{neumann}
von Neumann, J. (1932). \emph{Mathematical Foundations of Quantum Mechanics}, New Jersey: Princeton University Press.
\end{thebibliography}
\end{document}